\documentstyle[psfig]{aa}
\def\ltsima{$\; \buildrel < \over \sim \;$}
\def\simlt{\lower.5ex\hbox{\ltsima}}
\def\gtsima{$\; \buildrel > \over \sim \;$}
\def\simgt{\lower.5ex\hbox{\gtsima}}
\def\gsimeq
{\hbox{\raise0.5ex\hbox{$>\lower1.06ex\hbox{$\kern-1.07em{\sim}$}$}}}
\def\lsimeq
{\hbox{\raise0.5ex\hbox{$<\lower1.06ex\hbox{$\kern-1.07em{\sim}$}$}}}
\def\pn{\par\noindent}

\begin{document}
  \thesaurus{03(13.25.2; 11.19.1; 11.09.1: Mkn3)}

  \title{BeppoSAX observations of Mkn 3: Piercing through the torus of a Seyfert 2 galaxy}

   \author{M. Cappi,
              \inst{1}
     \and      L. Bassani
              \inst{1}
     \and      A. Comastri
              \inst{2}
     \and      M. Guainazzi
              \inst{3}
     \and      T. Maccacaro
              \inst{4}
     \and      G. Malaguti
              \inst{1}
     \and      G. Matt
              \inst{5}
     \and      G.G.C. Palumbo
              \inst{6,1}
     \and      P. Blanco
              \inst{7}
     \and      M. Dadina	
              \inst{8}
     \and      D. Dal Fiume
              \inst{1}
     \and      G. Di Cocco
              \inst{1}
     \and      A.C. Fabian
              \inst{9}
     \and      F. Frontera
              \inst{1}
     \and      R. Maiolino
              \inst{10}
     \and      L. Piro
              \inst{11}
     \and      M. Trifoglio
              \inst{1}
     \and      N. Zhang
	      \inst{12}
}
   \offprints{M. Cappi (mcappi@tesre.bo.cnr.it)}

  \institute {
       {Istituto TeSRE-CNR, Via Gobetti 101, I-40129
               Bologna, Italy}
  \and {Osservatorio Astronomico di Bologna, Via Ranzani 1, I-40127 Bologna, Italy}
  \and {Astrophysics Division, SCD -ESA, ESTEC, Postbus 299, NL-2200 AG
Noordwijk, The Netherlands}
  \and {Osservatorio Astronomico di Brera, Via Brera 28, I-20121 Milano, Italy}
  \and {Dipartimento di Fisica,Universita' di RomaIII, Via della Vasca Navale 
84, I-00146 Roma, Italy}
  \and {Dipartimento di Astronomia, Universita` di Bologna, Via Ranzani 1, 
I-40127 Bologna, Italy}
  \and {UCSD, San Diego Ca, USA}
  \and {BeppoSAX S.D.C., ASI, Via Corcolle 19, I-00131 Rome, Italy}
  \and {Institute of Astronomy, Cambridge University, Madingley Road,
Cambridge CB3 0HA,UK}
  \and {Osservatorio Astrofisico di Arcetri, Via L.E. Fermi 5, I-5015 
Firenze, Italy}
  \and {Istituto di Astrofisica Spaziale, Via Del Fosso del Cavaliere,
I-001333 Roma, Italy}
  \and {University Space Research Association, Huntsville-Al, USA}
}
   \date{Received / Accepted }

\titlerunning{BeppoSAX Observations of Mkn 3}
\authorrunning{M. Cappi et al.}

   \maketitle

   \begin{abstract}

A new $BeppoSAX$ broad-band (0.6--150 keV) spectrum of the Seyfert 2 galaxy 
Mkn 3 is presented. The spectrum provides a 
direct measurement of a large, neutral column of gas with $N_{\rm H}$ 
$\sim$ 10$^{24}$ cm$^{-2}$ in the source direction. The source, 
as bright as 3C 273 above 10 keV, has a steep ($\Gamma$ $\sim$ 1.8) 
spectrum without any evidence of a high-energy cutoff up to at least 150 keV.
At lower energies, the data are best modeled with the addition of 
an unabsorbed 
reflection component. Combining these data with previous $Ginga$ and 
$ASCA$ observations, the Fe K$_{\alpha}$ and reflection continuum indicate that 
the reprocessed emission is responding slower than the intrinsic 
continuum variations suggesting a size of the reprocessor $\gsimeq$ 2 pc. 
Identifying such a reprocessor with a (close to edge-on) obscuring torus, 
the overall result fits well into unified models since, presumably, 
one can interpret the strong absorption as due to transmission 
through the rim of the torus and the unabsorbed (directly viewed) 
reflection component as due to reprocessing from the torus inner surface.

 \end{abstract}

%

\section{Introduction}


It has long been recognized that the hard X-ray spectra of Seyfert galaxies 
(at E $\simgt$ 20 keV, photo-electric absorption becomes 
inefficient) are a powerful tool for extracting information on the intrinsic 
source emission properties (i.e. luminosity and spectral shape).
Such measurements are fundamental for the understanding of the 
emission mechanisms operating in these objects and, since they allow a direct comparison between different classes of AGNs (e.g. Seyfert 1 versus Seyfert 2 galaxies), 
for testing unified models (Antonucci 1993).
At lower energies ($\sim$ 0.1--20 keV), one can obtain 
information on column densities, ionization and abundances of the surrounding 
matter. This issue is important in testing the geometry and in particular the 
existence of molecular tori with 
$N_{\rm H}$ $\gsimeq$ 10$^{24}$ cm$^{-2}$ around Seyfert galaxies whose presence is 
essential for AGN unified models and synthesis models of the X-ray background.

In the few $OSSE$ data available for Seyfert galaxies, the steep high-energy spectra, 
the evidence of a high-energy cutoff (e.g. Maisack et al. 1993, Zdziarski et al. 1995, 
Grandi et al. 1998), 
and the absence of an annihilation line favour thermal Comptonization models 
(Haardt 1997) in contrast with the predictions from non-thermal pair models (see Svensson 
1994 for a review). More precise measurements by $BeppoSAX$ have confirmed the 
presence of high-energy cutoffs in some 
Seyfert 1 galaxies (Piro et al. 1999) though only lower-limits have been found 
in others (Perola et al. 1999).
To date, however, precise measurements are still lacking for Seyfert 2 galaxies with 
only a few attempted (Zdziarski et al. 1995, Bassani et al. 1995, 
Weaver et al. 1998), and, as such, fundamental to be performed.


Mkn 3 (z=0.0135) is one of the small sample of Seyfert 2 galaxies which show 
broad emission lines in polarized light suggesting the presence of a "hidden 
Seyfert 1 nucleus" (e.g., Miller \& Goodrich 1990 and Tran 1995). Other evidence
in favor of heavy obscuration are: the discovery of a biconical 
extended narrow line region (Pogge \& De Robertis 1993), 
the low flux of ionizing photons inferred from the directly observed UV 
continuum compared 
to the ionizing photons required to produce the observed H$\beta$ emission 
(Haniff et al. 1988, Wilson et al. 1988), the 
low L$_{\rm X}$/L$_{\rm [OIII]}$ ratio (0.14; Bassani et al. 1998) and the observation
of heavy X-ray obscuration. Indeed, {\it Ginga} data have
shown that the spectrum of Mkn 3 is flat ($\Gamma$ $\sim$ 1.3), absorbed 
by a column density of N$_{\rm H}$ $\sim$ 6 $\times$ 10$^{23}$ cm$^{-2}$ and has a
 strong iron line with equivalent width (EW) $\sim$ 500 eV (Awaki et al. 1991).
Measurements at low X-ray energies
revealed the presence of a soft excess and also indicated
the presence of prominent soft X-ray emission lines (Kruper et al. 1990, 
Turner et al. 1993, Iwasawa et al. 1994 (I94 hereinafter)). 
The {\it ASCA} observation revealed that the Fe K$_{\alpha}$ line emission 
decreased by a factor of 3 in response to a flux decline 
by a factor of 6 (I94). {\it ASCA} also resolved
the iron line in a 6.4 keV component of EW $\sim$ 900 eV and a 6.7 keV
one of EW $\sim$ 190 eV (I94). 
The analysis of the same data set
alone, or in combination with measurements from other instruments,
indicated the difficulty encountered in interpreting
unequivocally the 2-10 keV spectrum (Turner et al. 1997a,b and Griffiths et al. 1998, 
G98 hereinafter). It also highlighted the need for a broad-band
coverage to better interpret the observed emission.
At high energies, Mkn 3 was detected by $OSSE$ 
for the first time in March 1994 at a flux level of $\sim$ 2.8 
$\times$ 10$^{-11}$ erg cm$^{-2}$ s$^{-1}$
in the 50--150 keV energy band (Johnson, private communication).

Here we present the observation of Mkn 3 
by $BeppoSAX$. The results obtained highlight the potentialities of using $BeppoSAX$ in 
broad-band X-ray spectroscopy studies of active galactic 
nuclei, in particular in the case of highly absorbed sources like Seyfert 2 galaxies.
Throughout the analysis, we use H$_{\circ}$= 50 km s$^{-1}$ Mpc$^{-1}$ and 
q$_{\circ}$=0.

\section{Observations and data reduction}

The $BeppoSAX$ narrow field instruments consist of one low-energy concentration spectrometer 
(LECS, Parmar et al. 1997), three medium energy concentrator spectrometers 
(MECS, Boella et al. 1997), a high pressure gas scintillation proportional 
counter (HPGSPC, Manzo et al. 1997), and a phoswich detector system (PDS, Frontera et 
al. 1997) covering the 0.1--10 keV, 1.3--10 keV, 4--120 keV and 13--300 keV 
energy ranges, respectively.
HPGSPC data will not be considered in the present paper since the source is too 
faint for a correct background subtraction and we restricted the spectral analysis to 
the 0.1--4.5 keV and 1.5--10 keV energy bands for the LECS and MECS, respectively, 
where the latest released (September 1997) response matrices are best calibrated.
The LECS data were ignored below 0.6 keV, i.e. below 
the lowest energy at which Mkn 3 is detected at the 3 $\sigma$ level.

Mkn 3 was observed for about two days using all the above instruments during the 
period April 16--18, 1997. 
The total effective exposure was 39433 s in the LECS, 113690 s in the 
MECS and 51386 s in the PDS.
Light-curves and spectra were extracted (using the ftools v4.0 software package) 
from within a region of $\sim$ 4$^{\prime}$ radius centered on the source for both LECS 
and MECS instruments. 
Standard blank-sky files provided by the $BeppoSAX$ Science Data center (SDC) were used 
for the background subtraction.
Similar results were obtained using the background spectra extracted 
directly from the instruments fields of view (FOVs).
The 3 MECS instruments gave consistent timing and spectral results if taken 
individually and were, thus, added together. 
In the LECS and MECS FOVs, we also detected emission in the direction of the 
BL Lac object MS 0607.9+7108 about 6-7$^{\prime}$ NW of Mkn 3, that certainly 
lies also in the PDS FOV (which has a triangular response with 
FWHM of $\sim$ 1.3$^\circ$, Frontera et al. 1997). 
Although the statistics are too poor for a detailed spectral analysis of this source,
we can, however, exclude the possibility of contamination for the present analysis
because a) it is weak (about 1/100 the 2-10 keV flux of Mkn 3) and 
b) its spectrum is steep as indicated by the fact that its detection is clear in the 
LECS FOV but marginal in the MECS FOV. Similar considerations were also reported 
by I94 and G98.
Searches in high-energy catalogues reveal no other likely contaminating source within 
the 1.3$^\circ$ FOV of the PDS.

In total, the source count-rates were (6.8$\pm$0.5) $\times$ 10$^{-3}$ cts s$^{-1}$ 
in the LECS and 
(6.95$\pm$0.08) $\times$ 10$^{-2}$ cts s$^{-1}$ in the MECS, the background contributing 
$\sim$ 20\% at 6 keV and $\sim$ 7\% at 8 keV, respectively.
No significant variability was found within the statistical fluctuations of $\sim$ 
15 \% (MECS data), thus all data were accumulated over the whole observation.
The PDS data reduction was performed using both XAS (v.2.0, Chiappetti \& Dal Fiume 1997)
and SAXDAS (v.1.3.0.) software packages and yielded consistent results.
In the following, we'll refer to the results obtained with the XAS package.
The source count-rate in the PDS was 1.16 $\pm$ 0.02 cts s$^{-1}$ which 
corresponds to a total of $\sim$ 50 $\sigma$ detection in the 13-150 keV energy range
and $\sim$ 7 $\sigma$ between 100-150 keV. No significant variability was found 
but we cannot exclude variations up to a factor $\sim$ 2 (corresponding to the 
statistical fluctuations) within the entire observation.

For the spectral analysis, the LECS and MECS data were rebinned such as to sample the 
instrument resolution with a number of 1 and 5 channels per energy bin, 
respectively (about $\simgt$ 20 and 50 cts bin$^{-1}$). The PDS data were grouped 
logarithmically between 13-200 keV, with a S/N ratio $\simgt$ 3 per bin.
The spectral analysis was performed using version 10.00 of the XSPEC program 
(Arnaud 1996).


\section{Spectral analysis}

\subsection{The PDS high-energy spectrum}

A simple power-law fit (with $\Gamma$ $\simeq$ 1.61 $\pm$ 0.05; where 
N$_{E} \propto E^{-\Gamma}$) over the entire 
energy range ($\sim$ 13--200 keV) of the PDS data, is clearly unacceptable because
below $\sim$ 20 keV the data are systematically lower than the model, yielding a poor 
fit with a $\chi^{2}$ = 47 for 28 d.o.f. ($\chi^{2}_{\rm red}$=1.68, rejected at 99.8\% 
level in terms of $\chi^{2}$ statistics).
With absorption added to the model, the fit becomes acceptable with: 
$\Gamma$ $\simeq$ 1.79 $\pm$ 0.12, $N_{\rm H}$ $\simeq$ (1.54 $\pm$ 0.70) $\times$ 
10$^{24}$ cm$^{-2}$ and $\chi^{2}$ = 32.28 for 27 d.o.f. ($\chi^{2}_{\rm red}$=1.19).
Here and below, errors are at 90\% confidence for one interesting parameter 
unless otherwise stated. 
Best-fit spectrum, residuals and $\chi^{2}$ contour plots in the parameter space 
$N_{\rm H}$--$\Gamma$ are shown in Fig. 1.
If a 10\% systematic error is added to the data below 20 keV to account for residual 
calibration uncertainties at those energies (Dal Fiume 1998, private communication), 
we find somewhat larger 
($\sim$ 20\%) errors on $\Gamma$ and $N_{\rm H}$ but the derived 
best-fit values remain unchanged.
The observed (absorbed) 13--150 keV flux is about 1.47 $\times$ 10$^{-10}$ 
ergs cm$^{-2}$ s$^{-1}$, 
which makes Mkn 3 one of the brightest known AGNs in the hard X-ray energy range, 
as bright as 3C 273 (Haardt et al. 1998) and the buried Seyfert 2 galaxy 
NGC 4945 (Done et al. 1996).
Comparison with the $OSSE$ observations (Sect. 1) indicates that the source 
varied by at least a factor 3 in the 50--150 keV energy range.

\begin{figure}[htb]
\psfig{file=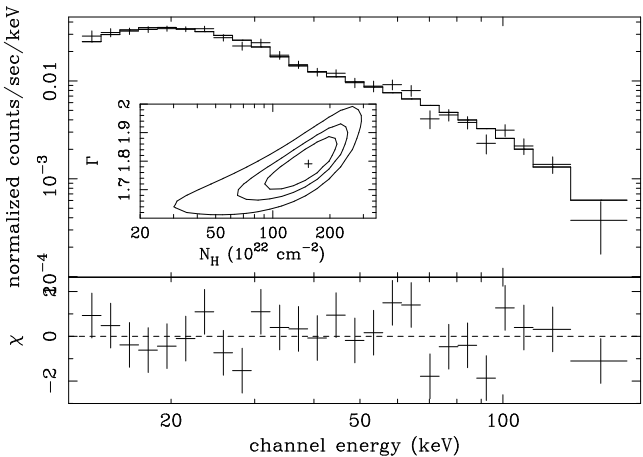,width=8.5cm,height=6cm,angle=-90}
\caption{Background subtracted PDS (13-150~keV) spectrum and residuals. Note that each 
data point has a S/N $\gsimeq$ 3. Inserted figures 
shows the $\chi^2$ contour plots in the $N_{\rm H}$-$\Gamma$ parameter space with 
solid line contours indicating the 68\%, 90\% and 99\% confidence limits.}
\end{figure}



\subsection{The 0.6--150 keV broad-band spectrum: baseline model}

Given the source spectral complexity (in particular below 10 keV), we addressed 
separately, in Appendix A, the study of the data below 10 keV {\it alone}. This 
allows us to compare {\it BeppoSAX} results to the ones previously reported in the 
literature, and to illustrate the ambiguities encountered in interpreting the spectrum 
of Mkn 3 over such a limited energy range.
In the following, the results obtained for the overall broad-band spectrum 
are presented.

The data sets from LECS (0.6--4.5 keV), MECS (2.5--10 keV) and PDS (13--150 keV) 
detectors were fitted simultaneously. To allow for differences in the absolute 
flux calibration of the individual detectors, the normalization of different 
instruments was allowed to vary within 20\% of the fiducial 
values of $A_{\rm LECS}$/$A_{\rm MECS}$ = 0.65 and $A_{\rm PDS}$/$A_{\rm MECS}$ = 
0.85 (e.g., Haardt et al. 1998).
Unless differently stated, the ratios obtained from the fits were all within 
$\sim$ 5\% of these fiducial values.
A column density $N_{\rm HGal}$ = 8.46 $\times$ 10$^{20}$ cm$^{-2}$ due 
to absorption by the Galaxy (Dickey \& Lockman 1990), was included in 
all models used in the following.

In Fig. 2, we show the residuals obtained between 0.6 and 150 keV for 
a fit with a power-law model plus Galactic absorption. 
The fit (where $\Gamma$ $\simeq$ 0.4)
clearly illustrates the complex spectral structure below 10 keV (see also Appendix A) and the 
strong and sharp ``rise and fall'' of the spectrum above 10 keV.

\begin{figure}[htb]
\psfig{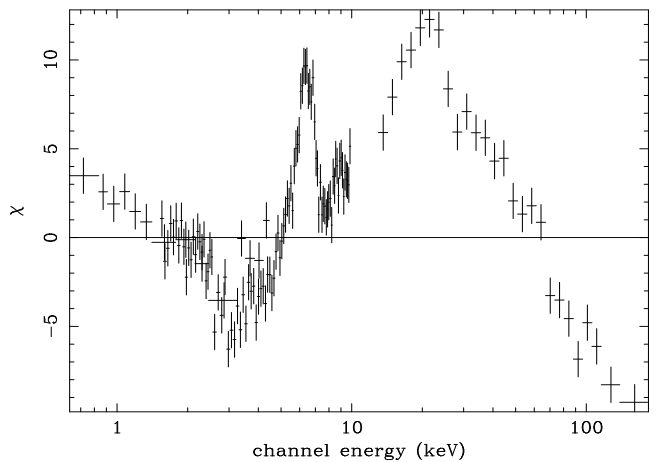}
\caption{Residuals (in units of standard deviation) 
of the LECS, MECS and PDS broad-band spectrum for a model consisting 
of a single power-law with $\Gamma$ $\sim$ 0.45. This figure illustrates the soft 
excess below $\sim$ 3 keV, the strong Fe K$_{\alpha}$ line at $\sim$ 6.4 keV and the sharp rise and fall 
of the spectrum at $E \gsimeq$ 10 keV.}
\end{figure}

Because of the presence of the very strong Fe K$_{\alpha}$ line, we first attempted to fit 
the overall broad-band spectrum with a pure reflection model
\footnote{The reflection 
model used is the $pexrav$ model in XSPEC which is an exponentially cutoff power-law 
spectrum reflected from a plane of neutral, optically thick, material 
(Magdziarz \& Zdziarski 1995). Inclination and abundances were fixed 
to the standard values, $\sim$ 60$^\circ$ and 1, respectively. The cutoff energy 
of the primary power-law spectrum ($E_c$) were frozen at 1000 keV, unless otherwise 
stated.} 
plus a steep soft power-law continuum that accounts for the low-energy emission 
below $\sim$ 3 keV.
Such a model is however ruled out ($\chi_{red}^2 \simeq 2$) since it falls 
short of the data between 8-20 keV for any value of photon index, cutoff, 
inclination and ionization state of the reflecting material.  
This clearly indicates that the sharp rise and fall observed above $\sim$ 10 keV cannot 
be explained {\it only} by the presence of an unabsorbed reflection component 
but requires, instead, 
the presence of a direct, strongly absorbed power-law component as modeled in Sect. 3.1.

In Appendix A, we show that a model consisting of a soft power-law continuum, an iron line 
plus a strongly absorbed hard power-law continuum (as initially proposed by 
I94) gives an acceptable description of the data below 10 keV. 
In the broad-band spectrum, however, we find that such model is ruled out 
($\chi_{red}^2$ $\simeq$ 1.4, Table 1) since it falls short of the data by a 
factor of about 2 in the PDS energy range.
The continuum below the line (from about 3 to 8 keV) 
is extremely flat, thus a simple transmission model from cold matter (even with $N_{\rm H}$ 
$\simeq$ 7--10 $\times$ 10$^{23}$ cm$^{-2}$) cannot {\it simultaneously} fit the low and 
high energy bands.
Similar results are obtained if one considers an ionized absorber instead 
of a neutral one. 

\begin{table*}[htb]
\caption{Fits of the Broad Band Spectrum}
\begin{center}
\begin{tabular}{ccccccccccccc}
\hline
\hline
\multicolumn{1}{c}{{\scriptsize Continuum Model$^{a}$}} &
\multicolumn{1}{c}{$\Gamma_s$} &
\multicolumn{1}{c}{{\scriptsize$A_s$/$A_h$}} &
\multicolumn{1}{c}{$N_{\rm H}^b$} &
\multicolumn{1}{c}{$\Gamma_h$} &
\multicolumn{1}{c}{$R$} &
\multicolumn{1}{c}{$E_{K_{\alpha}}^c$} &
\multicolumn{1}{c}{$\sigma_{K_{\alpha}}^d$} &
\multicolumn{1}{c}{$A_{K_{\alpha}}^e$} &
\multicolumn{1}{c}{{$\chi^{2}_{red}$/$\chi^{2}$/d.o.f.}} \\
\multicolumn{1}{c}{} &
\multicolumn{1}{c}{} &
\multicolumn{1}{c}{$\times$ 10$^{-2}$} &
\multicolumn{1}{c}{} &
\multicolumn{1}{c}{} &
\multicolumn{1}{c}{} &
\multicolumn{1}{c}{} &
\multicolumn{1}{c}{} &
\multicolumn{1}{c}{} &
\multicolumn{1}{c}{} \\
\hline
{\scriptsize pl$_s$+abs.pl$_h$} & 1.02$^{+0.18}_{-0.14}$ & 1.1 & 109$^{+16}_{-10}$ & 1.75$^{+0.08}_{-0.11}$ & - & 6.49$^{+0.07}_{-0.07}$ & 0.23$^{+0.14}_{-0.11}$ & 4.9$^{+1.7}_{-1.3}$ & 1.41/186/132 \\
{\scriptsize baseline: pl$_s$+refl.+abs.pl$_h$} & 2.64$^{+0.41}_{-0.79}$ & 2.3 & 127$^{+24}_{-22}$ & 1.79$^{+0.06}_{-0.12}$ & 0.95$^{+0.12}_{-0.15}$ & 6.47$^{+0.09}_{-0.12}$ &0.22$^{+0.16}_{-0.15}$ & 4.6$^{+1.7}_{-1.3}$ & 0.93/121/131\\
\hline
\hline
\end{tabular}
\end{center}
\pn
$^{a}$ See text for descriptions. $^{b}$ in units of 10$^{22}$ cm$^{-2}$. $^{c}$ Line 
energy in the source rest-frame, in units of keV. $^{d}$ Line width, in units of keV. 
$^{e}$ {\it Observed} line intensity in units of 10$^{-5}$ photons cm$^{-2}$ s$^{-1}$.
\pn
Note: Intervals are at 90 \% confidence for 2 interesting parameters.
\end{table*}

A better solution is found, instead, with a basic description (called ``baseline'' model 
in the following) as first proposed by Turner et al. (1997b) that includes 
a soft power-law component, a strongly absorbed power-law component and an iron line plus 
an {\it unabsorbed} reflection$^{1}$ component.
Photon index and normalization of the primary power-law were assumed to be 
those of the hard, absorbed, power-law component. 
In this simplest form, this model adds only 1 free parameter ($R$), namely the 
relative amount of reflection compared to the directly-viewed primary power-law.
The case $R$=1 corresponds to a 2$\pi$ coverage as viewed from the 
X-ray source.
The best-fit parameters obtained from this baseline model are given in line 2 of 
Table 1.
The unfolded spectrum and data-model ratios are given in Fig. 3.

\begin{figure}[htb]
\psfig{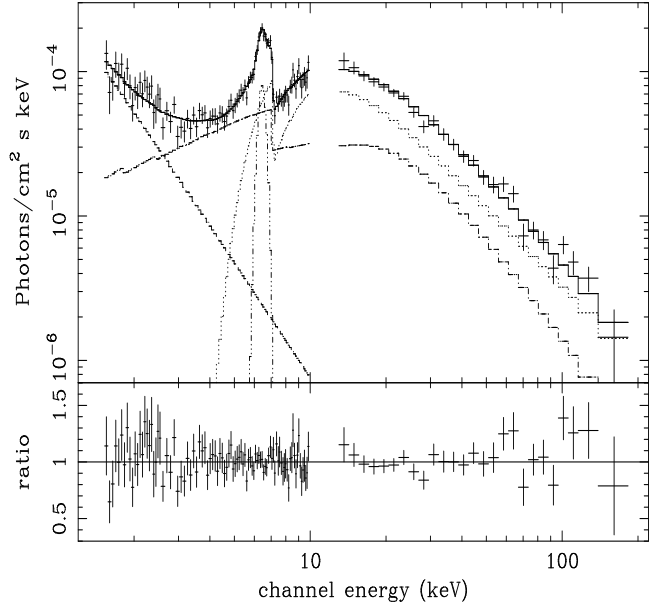}
\caption{Unfolded broad-band spectrum and baseline model obtained 
by fitting the LECS, MECS and PDS data. For clarity, only the MECS and PDS 
data are shown here.}
\end{figure}

\begin{figure}[htb]
\psfig{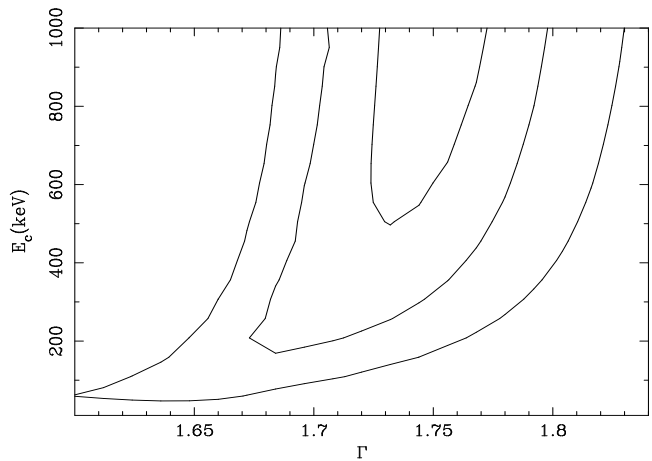}
\caption{$\chi^2$ contour plots in the $\Gamma$-$E_c$ parameter space. 
Solid line contours represent the 68\%, 90\% and 99\% confidence limits.}
\end{figure}

The cutoff energy ($E_c$) of the primary power-law 
continuum was then allowed to vary to search for a high-energy 
cutoff in the data.
The fit gives a lower limit of 
$E_c$ $\gsimeq$ 200 keV at 90\% confidence (see Fig. 4). However, given the 
10\% systematic uncertainty between the MECS and PDS normalizations 
(Cusumano et al. 1998), we estimated an additional $\sim$ 50 keV systematic error 
on this limit. 
It is stressed that it is difficult to obtain a very stringent limit 
on $E_c$ in this source because 
its intrinsic spectrum only emerges at E$\gsimeq$20 keV, where with the present data 
it is difficult to statistically sort out the parameters $R$, $N_{\rm H}$, $\Gamma$, 
and $E_c$.
However, looking at Fig. 3, we underline the clear absence of any cutoff 
at high energies up to about 150 keV. In conclusion, if a cutoff does exist, it occurs 
at energies above $\sim$ 150 keV.

%
%
%
%

The adopted baseline model (admittedly rather complex) is, however, 
too simple to describe correctly the overall complexity of the spectrum of Mkn 3.
In the following section, we discuss point by point the several ``deviations''
from the baseline model that are required by the data.
For completeness, more complex models, alternative to the proposed baseline model, 
are presented in Appendix B. Statistically, they represent viable alternative 
descriptions of the broad-band spectra. However, they were 
discarded on the basis of physical arguments and lack of simplicity 
(Occam's razor).


\subsection{``Optimizations'' of the baseline model}

First of all, in constructing the baseline model the absorption model used only 
considers photo-electric absorption and 
neglects Compton scattering which, however, becomes relevant for $N_{\rm H}$ $\gsimeq$ 
10$^{24}$ cm$^{-2}$ (Ghisellini et al. 1994, Yaqoob 1997). 
To take this effect into account, we used $plcabs$ in XSPEC which describes the X-ray transmission 
through a uniform, spherical distribution of matter, correctly taking into account 
Compton scattering (Yaqoob 1997).
As expected, this model gives a column density of $\sim$ 1.1 $\times$ 10$^{24}$ cm$^{-2}$,
slightly lower (though still consistent within the errors) 
than $\sim$ 1.3 $\times$ 10$^{24}$ cm$^{-2}$ obtained with 
our baseline model. At equal column density, the 
model that includes Compton scattering is more effective (by $\sim$ 20--30 \% between 
6 and 30 keV) in reducing the transmitted 
spectrum. The fit with $plcabs$ is statistically slightly worse than the one obtained with 
our baseline model ($\Delta \chi^2$ $\simeq$ 6). A slightly steeper ($\Gamma$ $\simeq$ 1.83)
high-energy power-law is obtained which, however, does not significantly change the 
other best-fit parameters (including the line parameters).
The main consequences of this correction are thus in the calculations of the 
source luminosity that strongly depends on the $N_{\rm H}$ and $\Gamma$ values 
used. With the $plcabs$ model, the 0.1--150 keV luminosity is reduced to about 
2.3 $\times$ 10$^{44}$ ergs s$^{-1}$, compared to the value of 2.6$\times$ 
10$^{44}$ ergs s$^{-1}$ obtained with the baseline model.

Secondly, there are still residuals at energies between $\sim$ 2 and 4 keV (Fig. 3), 
namely a broad absorption structure at $\sim$ 3 keV and some excess emission at 
$\sim$ 4 keV. Such excess counts indicate that the low-energy spectrum is probably more complex than 
a single power-law continuum, in agreement with {\it ASCA} results (Iwasawa 1995 (I95 hereinafter), 
G98). As suggested by these authors, thermal emission 
and/or absorption plus emission from a hot photo-ionized medium could be relevant in 
Mkn 3. Unfortunately, compared to the {\it ASCA} results, the poorer statistics 
of the LECS+MECS spectrum below $\sim$ 4 keV does not allow a detailed modeling of 
these features. Alternatively, we note that such features could be due to fluorescence of 
elements lighter 
than iron expected from the reflection component (and not included in our model) 
as calculated by Reynolds et al. (1994). 
For example, the feature at $\sim$ 4 keV 
could possibly be identified as a Calcium K$\alpha$ fluorescence line emission (3.69 keV). 
In any case, the baseline best-fit parameters only weakly depend on the 
exact modeling of the soft component because variations by $\sim$ 50\% in the soft 
power-law slope affect $N_{\rm H}$, $\Gamma_{\rm hard}$ and A$_{K\alpha}$ by less than 
10\%, i.e. less than the statistical errors reported in Table 1.

Thirdly, the power-law continuum of our baseline model represents only a 
phenomenological model.
We thus tested a more physical and self-consistent model: the
Comptonization model ($tita_a$ in XSPEC) calculated by Hua \& Titarchuck (1995), 
that also includes reflection from optically thick gas.
This model basically replaces the underlying power-law continuum with 
an exponential cutoff of our baseline model by a thermal Comptonization spectrum 
with reflection (Magdziarz \& Zdziarski, 1995). 
The fundamental parameters of the model are the thermal gas optical depth ($\tau$) 
and its temperature ($kT$). 
We find that the data can be reasonably well modeled ($\chi^2_{red}$ = 0.99) by an intrinsic 
spectrum due to thermal Comptonization with $\tau~\simeq~0.5\pm~0.2$ and 
$kT~\gsimeq~110$ keV (at the 90\% confidence level for two interesting parameters). 

Fourthly, the baseline model requires a broad iron line whose modeling 
is addressed in detail in the following section.

\subsection{On the iron K line complex}

It is interesting to note that if the shape of the continuum below the line 
(in the energy interval $\sim$ 3--10 keV) is parameterized by a single power-law 
model (with $\Gamma$ = -0.5), 
the line residuals (Fig. 5) suggest a complex profile. It comprises a red-shifted 
wing down to about 5 keV, a narrow and strong component peaking at $\sim$ 
6.4 keV plus a clear wing on the high-energy side of the line which drops 
at $\sim$ 7 keV.
A continuum $\Gamma$ = -0.5 model is clearly unphysical and should 
rise doubts on the interpretation of the line profile obtained in this way.
However, such multi-peaked 
structure may suggest a physical origin in an accretion 
disk (for the red and blue wings) plus a superposition of a narrow line 
at $\sim$ 6.4 keV produced far from the black hole. Such conditions have been found 
in other Seyfert 2 galaxies often grouped under the name of Narrow Emission Line Galaxies 
(Weaver et al. 1997, Weaver \& Reynolds 1998).
However, contrary to what found by these authors in the sources they analyzed, 
in Sect. 3.1 and 3.2 we have shown that Mkn 3 certainly has a large absorption column 
which strongly affects the modeling of the underlying continuum between 3--10 keV.
Indeed if we parameterize the continuum as given by our baseline model, 
we obtain the residuals shown in Fig. 6.
Clearly, the red wing can be almost entirely explained by the combination of 
the reflection continuum and strong absorption. This illustrates the 
complexities of spectral modeling of emission lines when the 
underlying continuum is not well defined.
The intrinsic power of broad band spectroscopy in sorting out these 
complexities and the advantage of using $BeppoSAX$ in this case are clearly evidenced.

\begin{figure}[htb]
\psfig{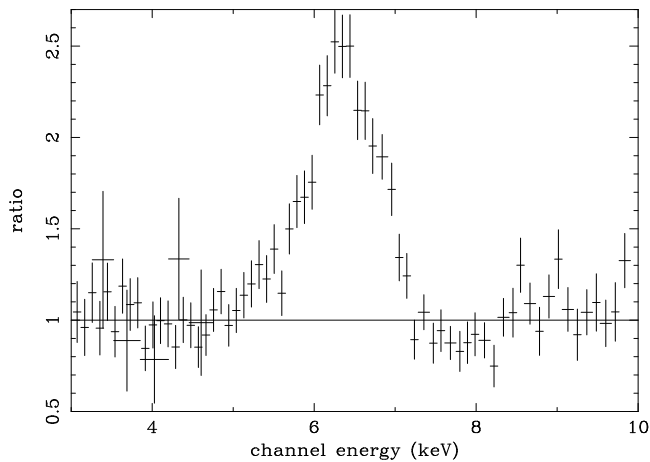}
\caption{Iron line feature obtained from fitting the LECS + MECS spectrum 
between 3--5 keV and 7--10 keV with a single power-law model ($\Gamma~\sim-0.5$).}
\end{figure}
\begin{figure}[htb]
\psfig{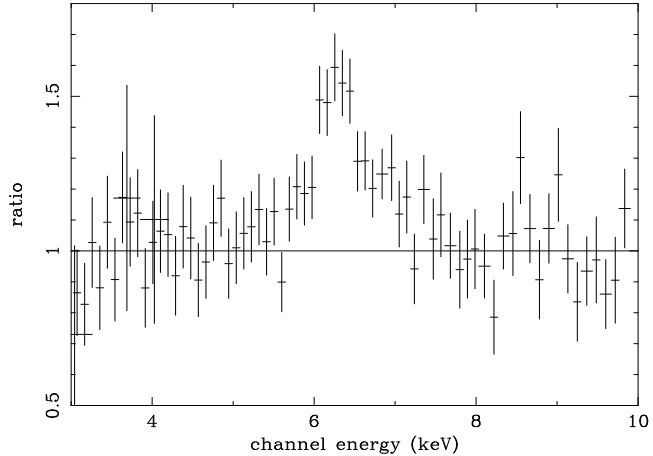}
\caption{Data to model ratios obtained from fitting the LECS + MECS + PDS spectrum 
between 0.8--5 keV and 7--100 keV with the baseline model (Table 2, line 4).}
\end{figure}

However, Fig. 6 shows that ``even'' with 
our baseline model, the line appears complex with a significant excess emission at 
energies between 6.4--7 keV.
Fitting the data with two narrow ($\sigma$ = 0 eV) Gaussian lines yields a significantly 
better fit ($\Delta \chi^2$ = 6, for one more free parameter). The fit 
gives E$_1$= 6.43$^{+0.08}_{-0.08}$ keV, 
I$_1$ = 3.5$^{+0.9}_{-1.0}$ $\times$ 10$^{-5}$ photons cm$^{-2}$ s$^{-1}$ (observed) for the neutral 
fluorescence iron line and E$_2$ = 7.00$^{+0.23}_{-0.24}$ keV, I$_2$ = 1.3$^{+0.9}_{-0.9}$ 
$\times$ 10$^{-5}$ photons cm$^{-2}$ s$^{-1}$ (observed) for the line at higher energies, respectively. 
These values are similar to the results previously 
found with {\it ASCA} (I94, G98, Netzer et al. 1998).
The observed intensity of the neutral line corresponds to an equivalent width of 
$\simeq$ 645 $\pm$ 180 eV with respect to the reflection component and 
$\simeq$ 650 $\pm$ 182 eV with respect to the absorbed power-law continuum. 
The observed intensity of the line at higher energies corresponds to an equivalent width of 
$\simeq$ 235 $\pm$ 160 eV, $\simeq$ 164 $\pm$ 113 eV and $\simeq$ 6.2 $\pm$ 4.3 keV 
with respect to the reflection component, the absorbed power-law continuum and 
the (soft) scattered continuum, respectively.

From Fig. 6, there is also an indication of a weak red-wing of the 6.4 keV 
iron line that could be attributed to Compton down-scattering in optically 
thick, cold matter as found in the Seyfert 2 galaxy NGC 1068 by Iwasawa et al. (1997). 
Theoretical models predict that this Compton shoulder should contribute with an 
intensity about one tenth that of the line core (Matt et al. 1991). 
Unfortunately, we cannot reach any firm conclusion about this 
because the feature is not statistically significant in the data.

\subsection{Long-term X-ray history of Mkn 3}

The {\it observed} 2--10 keV flux (calculated with our baseline model) is 
6.5 $\times$ 10$^{-12}$ ergs cm$^{-2}$ s$^{-1}$ which places Mkn 3 in an 
intermediate state between its brighter state observed by $Ginga$ in 1989 
(F$_{\rm 2-10 keV}$ $\sim$ 6--10 $\times$ 10$^{-12}$ ergs cm$^{-2}$ s$^{-1}$, 
depending on the adopted model: Awaki \& Koyama 1993, I95, G98) 
and the fainter state measured by $ASCA$ in 1993 (F$_{\rm 2-10 keV}$ $\sim$ 1.3--1.9 
$\times$ 10$^{-12}$ ergs cm$^{-2}$ s$^{-1}$, depending on the adopted model: 
I94, G98, Turner et al. 1997a).

Given the spectral complexity of Mkn 3, it is not straightforward to compare the 
present $BeppoSAX$ results with previous X-ray observations because any residual 
feature (in particular the iron line intensity and continuum flux) is a strong function 
of the 
model adopted for the underlying continuum (see Appendix A) and also depends inevitably 
on the instrumental sensitivity and resolution.
In order to provide a fair comparison, we thus re-analyzed the archival {\it ASCA} GIS 
and {\it Ginga} 
(upper-layer) data and compared it directly to the $BeppoSAX$ MECS data.
To reduce as much as possible the model-dependency in this comparison, we fitted each 
dataset only between 3--10 keV using a single power-law model with $\Gamma$ 
fixed to -0.25 (the average value obtained from the 3 instruments) and free normalization. 
A direct comparison between the three unfolded spectra is shown in Fig. 7. Inspection 
of the figure 
reveals that 
i) the continuum above the line (E $>$ 6.5 keV) decreased by about a factor of 4 between 
$Ginga$ and 
$ASCA$ and increased by about a factor 2 between $ASCA$ and $BeppoSAX$, 
ii) the $\sim$ 5--6 keV continuum and Fe K$_{\alpha}$ line intensity varied significantly less, 
by a factor of about 2 between $Ginga$ and $ASCA$, and by less than 50\% between 
$ASCA$ and $BeppoSAX$ and iii) the 3--5 keV continuum did not vary at all.
The slow responses of the Fe K$_{\alpha}$ line and 3--6 keV continuum to the flux variations clearly 
indicate that 
such variations cannot be due to pure transmission but need dilution by a constant component.
The argument, in turn, is 
giving support to the results of our spectral analysis (Sect. 3.2) which require 
inclusion of a Compton reflection component.
As a matter of fact, all observations are consistent with a variable direct component and 
associated (variable) transmitted Fe K$_{\alpha}$ line plus a constant reflection component and 
associated (constant) reprocessed Fe K$_{\alpha}$ line.
The non-varying (reprocessed) Fe K$_{\alpha}$ line indicates that it did not respond to continuum 
variations occured 7 years apart and places a rough lower limit of $\sim$ 2 pc from the 
central source 
on the location of the Fe K$_{\alpha}$ line emitter.
It should also be noted that the implication that the direct component of Mkn 3 varies with time 
is also supported by the observed increase (by a factor of $\sim$ 3) between the $OSSE$ and 
$BeppoSAX$ high energy fluxes (see Sect. 3.1).

\begin{figure}[htb]
\psfig{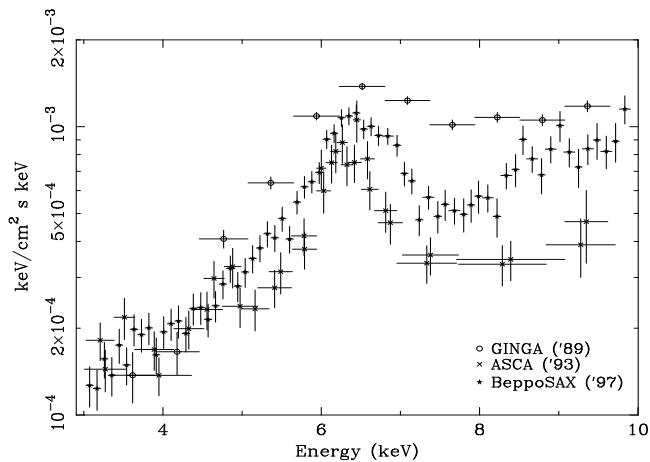}
\caption{Overplot of the $BeppoSAX$ (MECS), $ASCA$ (GIS) and $Ginga$ (top-layer) 
unfolded spectra. See text for details. 3--10 keV data fitted with a single (inverted) 
power-law with $\Gamma$=-0.25.}
\end{figure}

In the soft (E$<$ 3 keV) X-ray band, the $BeppoSAX$ flux of $\sim$ 7.6 $\times$ 
10$^{-13}$ ergs cm$^{-2}$ s$^{-1}$ between 0.6--3 keV (corresponding to a luminosity 
of $\sim$ 6 $\times$ 10$^{41}$ ergs s$^{-1}$) is comparable 
to previous measurements with the $Einstein$ IPC (Kruper et al. 1990), 
$BBXRT$ (Marshall et al. 
1992), $ROSAT$ PSPC (Turner et al. 1993) and $ASCA$ (I95). 
Therefore, as previously pointed out by I95, the lack 
of variability of the soft X-rays suggests that they originate from an extended region and 
immediately rules out a partial covering of the X-ray continuum which 
predicts that both hard and soft X-rays should vary simultaneously.
Mkn 3 does not show evidence for starburst contamination (Pogge \& De Robertis 1993) 
and Turner et al. (1997b) estimated a thermal emission in the 0.5--4.5 keV band 
lower than $\sim$ 6 $\times$ 10$^{40}$ ergs s$^{-1}$ on the basis of its 
far infra-red luminosity. Therefore the most probable explanation for the lack 
of variations is that the soft X-rays 
are dominated by scattering of the intrinsic continuum (see also G98 and Netzer et al. 1998).

\section{Discussion}

The above results have shown that the broad-band spectrum of Mkn 3 
is best described by the sum of a soft power-law, an {\it unabsorbed} 
reflection component, an iron line and a strongly absorbed hard power-law 
component.
The most remarkable result of the present analysis is that the hard (E$\gsimeq$ 20 keV) X-ray spectrum 
is steep with $\Gamma$ $\sim$ 1.8 and that there is no evidence of a spectral break 
in the data for energies up to at least $\sim$ 150 keV. 
The steep intrinsic spectrum is consistent with the canonical value found 
for Seyfert 1 galaxies (Nandra \& Pounds 1994, Perola et al. 1999). The 
lack of a cutoff in our data is consistent with the large average 
value ($E_c$ $\simeq$ 0.7$^{+2.0}_{-0.3}$ MeV) found for a sample of 
5 radio-quiet Seyfert 1 galaxies detected both by $OSSE$ and $Ginga$ (Gondek et al. 
1996, Zdziarski et al. 1997). It is also consistent with the lower-limits 
($E_c$ $\gsimeq$ 150 keV, Perola et al. 1999) and detections 
($E_c$ $\simeq$ 150--200 keV, Piro et al. 1999) obtained with $BeppoSAX$ 
for several Seyfert 1--1.5 galaxies.
Both measurements thus support a unified model of Seyfert galaxies.

The very high absorption column density measured here makes Mkn 3 very 
similar to other known buried Seyfert 2 galaxies: NGC 4945 
(Done et al. 1996), Circinus galaxy 
(Matt et al., in preparation) and possibly NGC 4941 (Salvati et al. 1997).
As pointed out by Salvati et al. (1997), a better knowledge of the number 
of such heavily absorbed sources is extremely important because it 
has direct consequences on synthesis models of the X-ray background 
(Comastri et al. 1995, Gilli et al. 1999).
In the 3--6 keV band, the spectrum of Mkn 3 requires a relevant contribution from 
reprocessed emission, namely an unabsorbed reflection component with associated 
iron emission line. The best-fit parameters obtained (R $\sim$ 0.95, E(Fe K$_{\alpha}$) $\sim$ 6.4 keV, 
and EW(Fe K$_{\alpha}$) $\sim$ 650 eV with respect to the reflection component) are comparable to 
the theoretical values expected for a cold reflector covering a solid angle of $\sim$ 2 $\pi$ 
at the source (George \& Fabian 1991, Matt et al. 1991). 
The long-term variability study suggests a distance of the reprocessor $\gsimeq$ 
7 light years (Sect. 3.5), that could thus be identified 
with an obscuring torus. Similar results have been found for example 
in NGC 2992 (Weaver et al. 1996), NGC 4151 (Piro et al. 1999) and NGC 4051 
(Guainazzi et al. 1998).
Moreover, a $\sim$ 2 $\pi$ covering of the reprocessor is also 
consistent with a torus interpretation, provided that its half-opening angle is 
$\sim$ 45\%.

In conclusion, both the strong absorption and unabsorbed reflection can be explained in the 
framework of unified models (i.e. assuming the existence of an optically thick torus). 
The absorption resulting from the transmission of the direct 
component through the rim of the torus. The reflection component (observed directly) 
resulting from reprocessing of the (same) direct component by the inner surface of the torus.

The origin of the second narrow line at higher energies is instead puzzling. Its energy 
(E $\sim$ 7.0 keV) and intensity (EW $\sim$ 235 $\pm$ 160 eV) are roughly consistent with 
fluorescence iron K$\beta$ emission from either the reflection component or the absorption 
component. Theoretical models would in fact predict (E $\sim$ 7.06 keV and EW $\sim$ 72 eV, 
assuming a K$\beta$/K$\alpha$ ratio of 1:9, Matt et al. 1996).
Alternatively, it could be interpreted as H- and/or He-like iron emission
produced by the scattered soft component itself. Indeed, as shown by Matt et al. 
(1996), very strong resonantly scattered H- and He-like lines 
(with EW between 2--4 keV with respect to the scattered component, 
depending on the material optical depth and the line being strongest in the 
optically thin regime) are expected if warm material is responsible 
for the scattering. Given the large uncertainties in our measurement of its 
equivalent width, especially when calculated with respect to the soft 
scattered component (Sect. 3.4), such possibility cannot be excluded. 
There could also be a contribution from both K$\beta$ fluorescence 
and resonant scattering.
Moreover, we cannot exclude that both the weak red and blue wings of the Fe K$_{\alpha}$ line (Fig. 6) 
are produced by an additional reflection component (with associated diskline) 
from an accretion disk, as commonly observed in Seyfert 1 galaxies (Nandra \& Pounds 1994, 
Perola et al. 1999) and, possibly, in some Narrow Emission Line Galaxies 
(Weaver \& Reynolds 1998). Reflection from a highly inclined relativistic accretion 
disk would indeed produce extra emission at energies below and above the 
transmitted 6.4 keV Fe K$_{\alpha}$ line.
In the case of Mkn 3, the effects of strong absorption and poor statistics hamper, 
however,  a detailed modeling of such additional component that will require the use of 
more sensitive instruments like $AXAF$ or $XMM$.


\section{Summary}

The main results derived from the present $BeppoSAX$ observation of Mkn 3 are the following:

- Fit of the data above $\simeq$ 8 keV require a strongly absorbed power-law, with 
$N_{\rm H}$ $\simeq$ 1.3 $\times$ 10$^{24}$ cm$^{-2}$.

- The intrinsic high-energy power-law 
continuum is steep ($\Gamma$ $\sim$ 1.8) and shows no high-energy cutoff up to 
$\sim$ 150 keV, similar to what is found in most Seyfert 1 galaxies.

- Having argued that a second absorbed power-law (i.e. a dual-absorber) is unlikely 
in this source, the data between $\sim$ 2-10 keV require an unabsorbed 
reflection component to make the flat spectrum underlying the iron line.

- From the data fit and the long-term X-ray history of the source, the (neutral) iron line intensity is 
consistent with being produced in an obscuring torus placed at a distance of 
$\gsimeq$ 2 pc from the source partly by transmission through its rim and partly by
reflection over its inner surface.

- Iron line emission is also detected at $\sim$ 7.0 keV. It can be interpreted 
either as K$_{\alpha}$ emission associated to the neutral reflection component and/or 
to the absorption, either as H- and He-like Fe emission associated to the ionized soft 
scattered component, or a combination of the two.
It could also be produced by an additional reflection component from a highly 
inclined relativistic accretion disk.
 
Overall, these results fit well into unified models scenarios of Seyfert galaxies 
and highlight
the potentialities of using $BeppoSAX$ in broad-band X-ray spectroscopy studies of 
buried sources like Seyfert 2 galaxies.

\begin{acknowledgements}
We thank the $BeppoSAX$ Team for contributing to the operations of the satellite and 
for continuous maintenance of the software.
This research has made use of SAXDAS linearized and cleaned event
files produced at the $BeppoSAX$ Science Data Center. 
Financial support from the Italian Space Agency is acknowledged.
We thank D.A. Smith who kindly provided us the $Ginga$ spectrum of Mkn 3 
and an anonymous referee for detailed and constructive remarks.

\end{acknowledgements}

\begin{figure}[htb]
\psfig{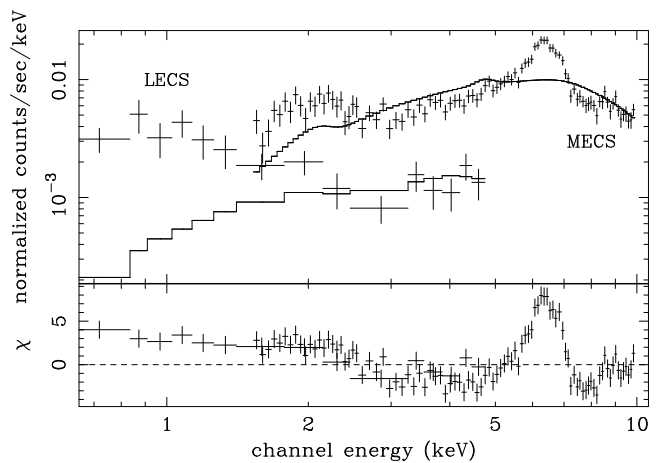}
\caption{LECS and MECS spectra and residuals with a continuum model consisting in 
a single power-law model ($\Gamma$ $\sim$ -0.5) absorbed by the Galactic absorption 
$N_{\rm Hgal}$=8.46 $\times$ 10$^{20}$cm$^{-2}$.}
\end{figure}

\begin{figure}[htb]
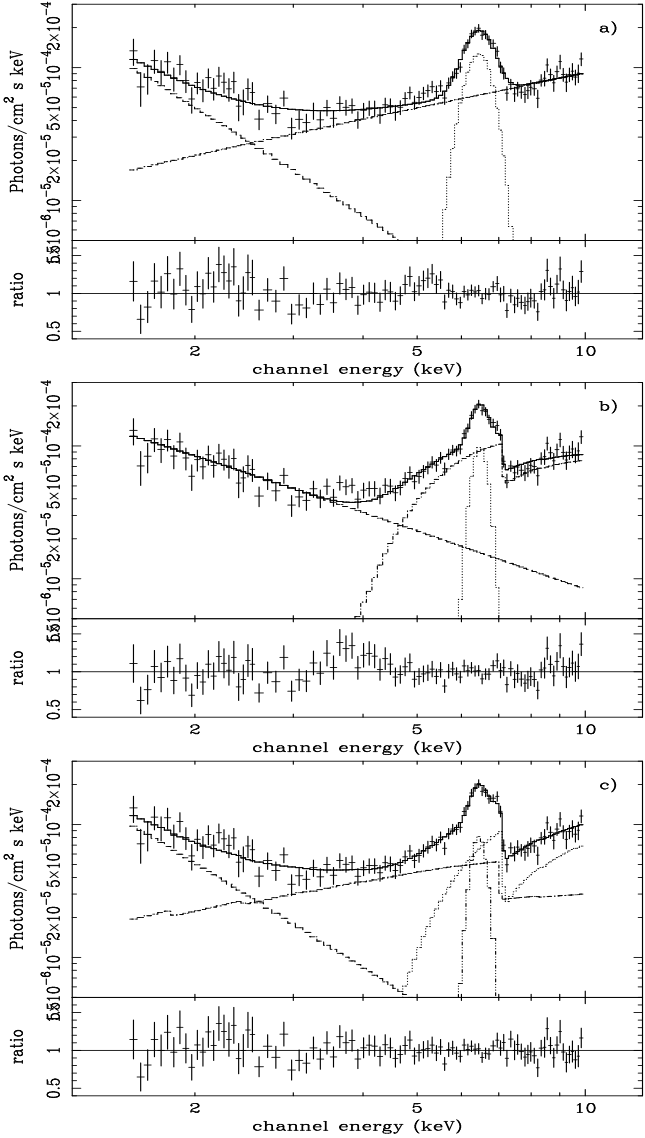

\psfig{file=./mcappi_f9a.ps,width=8.5cm,height=5cm,angle=-90}
\psfig{file=./mcappi_f9b.ps,width=8.5cm,height=5cm,angle=-90}
\psfig{file=./mcappi_f9c.ps,width=8.5cm,height=5cm,angle=-90}
\caption{Unfolded spectra with best-fit models as obtained from the fits of the LECS+MECS 
data. Pannels a), b) and c) correspond to lines 1, 2 and 3 of Table A.1, respectively.
For clarity, only the MECS data have been plotted here.}
\end{figure}

\begin{figure}[htb]
\psfig{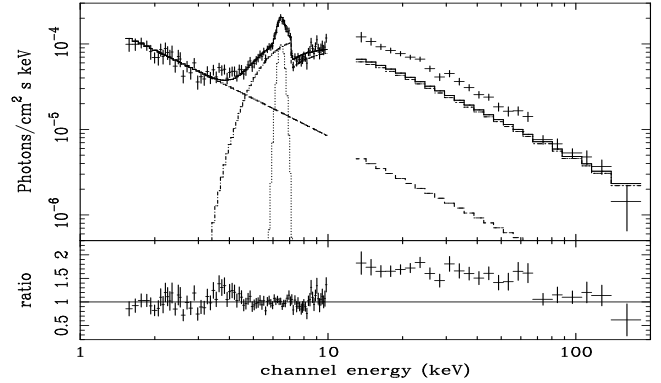}
\caption{Unfolded spectrum with the best-fit model b) as obtained from the fit of 
the LECS+MECS data and with the PDS data added subsequently.}
\end{figure}

\begin{appendix}

{\bf APPENDIX A:} {\it Comparison with results and ``ambiguities'' 
from the literature for the data below 10 keV}

To allow a direct comparison of the {\it BeppoSAX} results with the results reported 
in the literature (Sect. 1), we restrict here the analysis to the data below 10 keV. 
In agreement with previous findings (G98 and ref. therein), 
the spectrum of Mkn 3 below 10 keV is very complex, with prominent soft 
excess emission below 3 keV and strong iron 
line emission above a flat underlying 3--10 keV continuum. This is clearly illustrated by 
the residuals shown in Fig. 8 obtained by simultaneously fitting the LECS and MECS data 
with a single power-law model, with an extremely flat best-fit 
photon index $\Gamma$ $\sim -0.5$.

The three basic descriptions for the continuum shown here (Table A.1) comprise 
{\bf a)} a double power-law model; {\bf b)} a soft power-law plus a hard, absorbed, power-law 
model; and {\bf c)} model b) plus an 
{\it unabsorbed} pure reflection continuum model.
A Gaussian emission line is added in all fits. 
The resulting best-fit parameters are given in Table A.1.

We find that model c gives the best description of the data at energies lower than $10$ keV, 
with a $\Delta \chi^2 \sim 35$ compared with model b for one additional free parameter, 
in agreement with Turner et al. (1997b).
However, it is difficult to clearly prefer one model 
to the other. Inclusion in model b of extra emission/absorption 
features in the soft and/or hard component (rather plausible if the scattering gas and/or 
absorbing material is ionized) could account for most of the remaining residuals of 
the fit (i.e, the 3--5 keV bump, iron edge structure and excess emission above 
$\sim$ 7 keV, as shown in Fig. 9b). 
This ambiguity is similar to the one 
resulted from the {\it ASCA} data that led to different parameterizations and  
interpretation of the same data by different authors (I94, Turner et al. 1997b, G98).
However, we have shown in Sect. 3.2 how such an ambiguity can be solved thanks 
primarily to the {\it BeppoSAX} detection of the high-energy ($E~\gsimeq~10$ keV) emission.
This is also illustrated in Fig. 10 which shows that an extrapolation of 
model b) to higher energies falls short of the PDS data, thus requiring a more 
complex model with larger absorption.

Because of the spectral complexity, we also find that any residual feature 
is a strong function of the adopted model for the underlying continuum.
For example, from the ratios shown in Figs. 9a, 9b and 9c there is some 
evidence of a broad absorption structure between $\sim$ 7--8 keV that, at first glance, 
could be ascribed to absorption by ionized material (G98). 
However, a comparison of 
the different figures shows that the feature is significant within model a and 
model b but is only marginal in model c.
Indeed, in our baseline model, such an edge is entirely accounted for 
($\tau \lsimeq$ 0.5 if an extra absorption edge is added 
at energies between 7--8 keV) by the combination of the deep Fe K edge of the large 
absorption column density plus the broad Fe K edge produced by the reflection component.

Moreover, as shown in Table A.1, the same data give differences of about 
a factor of two in the line width (and thus line intensity) depending on the 
adopted continuum model. Even for equal width (compare lines in pannels 
b and c of Fig. 9), a different continuum modeling can result in a $\sim$ 20\% 
difference in the 
{\it observed} line intensity. 

In conclusion, the above results highlight the need of broad-band spectroscopy for such 
complex sources in order to reach confidence that the continuum emission and spectral 
features are properly modeled.


{\bf APPENDIX B:} {\it Alternative models to our baseline model}

More complex models have also been fitted to the 
\pn
$BeppoSAX$ broad-band spectrum, in particular: 
a {\it neutral} dual absorber model and a dual absorber model with a mixture 
of neutral and ionized matter.
The basic idea is to explain the flat $\sim$ 3--10 keV underlying 
continuum without using an {\it unabsorbed} reflection component (as proposed
in our baseline model, Sect. 3.2) but with the addition of a second absorbed power-law 
component.
The former model has been extensively used in the literature
to explain complex absorption and/or flat 2--10 keV spectra of several Seyfert galaxies 
(EXO 055620-3820.2, Turner et al. 1996; NGC 4151, Weaver et al. 1994; NGC 5252, Cappi et 
al. 1996; NGC 2110, Hayashi et al. 1996, Malaguti et al., in preparation; NGC 7172, 
Guainazzi et al. 1997; IRAS 04575-7537, Vignali et al. 1998).
The latter model has been first proposed by G98 for Mkn 3 based 
primarily on the {\it Ginga} detection of an ionized Fe K edge at $E \sim$ 8 keV.
The results obtained with these models are given in Table B.1. The ionized absorber model 
used here is the {\it absori} model in XSPEC (Done et al. 1992) with a temperature 
fixed to 10$^6$ K, as in G98 (none of the following conclusions changed, though, for 
a temperature ranging between 10$^{4-6}$K).
We find that both models give acceptable fits of the broad band spectrum and are 
both statistically indistinguishable given our data alone.
None of them, however, gives a statistically better fit than the baseline model 
(Sect. 3.2, Table 1) considering the increase by 1 and 2 in the number of free parameters 
for the former and latter model, respectively. 
Given the poor statistics of the data at low energies, we did not attempt 
to fit a more physical (and more complex) model for the ionized absorber 
that would take into account emission/reflection from the absorbing gas or 
the possible presence of dust.

Moreover, the main problem with both these models is that, in order to interpret 
physically different absorbing columns along our line of sight, 
one is forced to assume either that i) the source of X-rays is not 
a point source as seen from the two absorbers {\it and} the two different columns 
cover different areas of the source or that ii) the lowest of the two column densities 
covers the whole source while the larger one only partially covers it.
The former geometry appears rather unlikely given the small dimensions generally 
inferred to the X-ray emitting regions in AGNs from variability arguments 
(e.g. Mushotzky et al. 1993).
The latter geometry could be more physical and fits well into the framework 
of Unified Models since one could identify the highest column density ($N_{\rm H}(1)$ in 
Table B.1) with the BLR clouds, partially covering the source, and the lowest 
column ($N_{\rm H}(2)$) with absorbing matter, 
eventually ionized, associated with the torus (or its rim) or with some matter in the outer 
zone of the galaxy (e.g. Hayashi et al. 1996, Vignali et al. 1998). 
However, our best-fit results give $N_{\rm H}(1)$ $\sim$ 1.2 $\times$ 10$^{24}$ cm$^{-2}$ 
and a covering fraction of 90\% (Table B.1, column 7) which are at least an order 
of magnitude larger than commonly believed for the BLR (respectively 10$^{22-23}$ 
cm$^{-2}$ and 5--30\%, Netzer \& Laor 1993, Kwan \& Krolik 1981).
Such complex models may, possibly, also find alternative interpretations in terms 
of different geometries involving complex scattering by matter at 
non-uniform density and/or with non-uniform coverage.

It should also be noted that the main reason G98 proposed 
an ionized absorption model was the presence of an ionized Fe K absorption edge
in the $Ginga$ data at $\sim$ 7--8 keV. 
However our MECS 7--10 keV data, with unprecedent statistics, show that 
such edge could be entirely accounted for by the 
combination of the reflection component and the larger column density 
(required by the PDS data) as given in our baseline model (Sect. 3.2).
In conclusion, we find no need from 
the present data to invoke more complex models and/or to invoke ionized absorption 
in our data.

\end{appendix}

\vfill\eject
\onecolumn

 
\begin{figure}[htb]
\vspace{-7cm}
\psfig{file=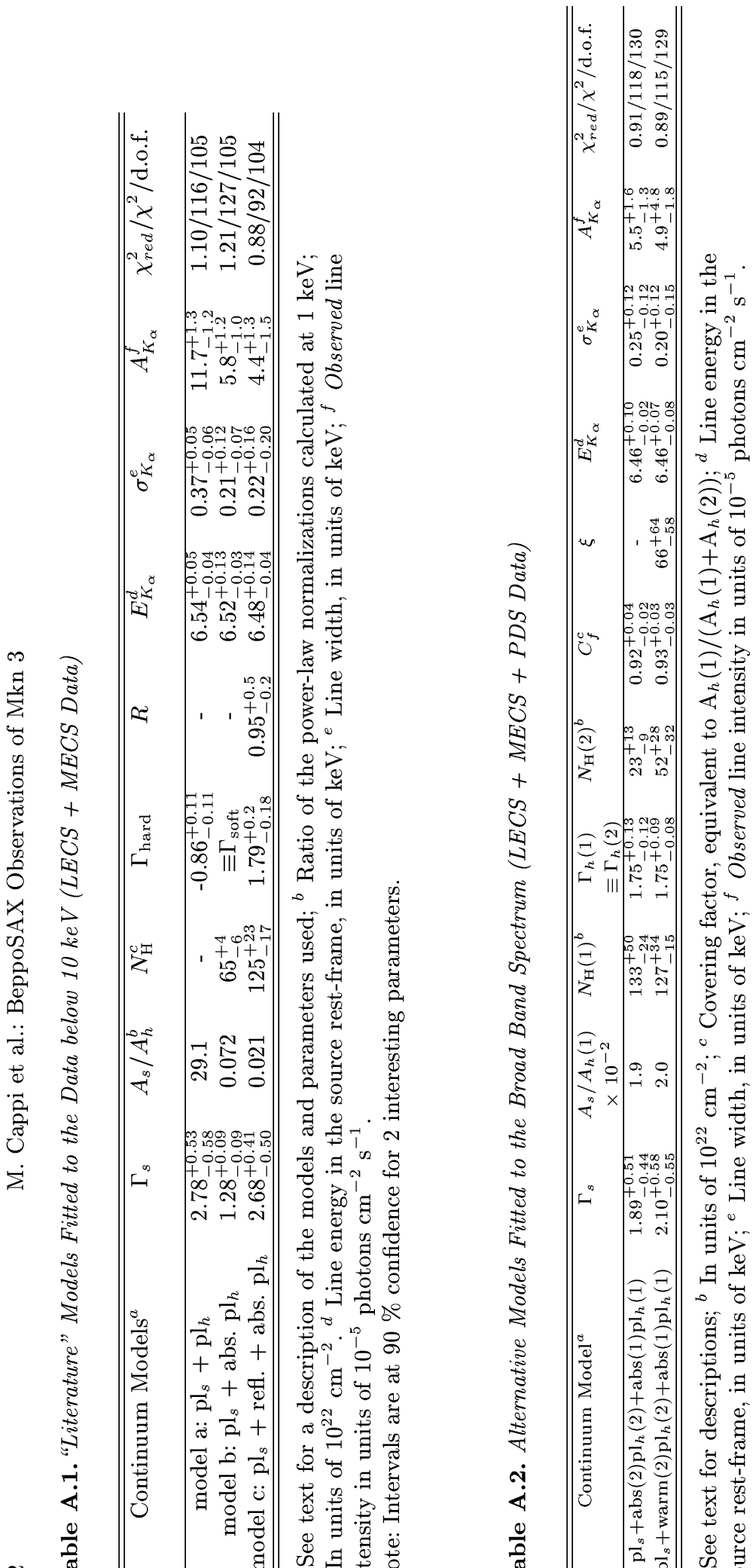,width=20cm,height=28cm,angle=180}
\end{figure}
\end{document}